\pdfoutput=1
\pdfmapfile{+classico.map} 
\newif\ifafour
\afourfalse 
\newif\iftypodisclaim 
\typodisclaimfalse

\documentclass[\ifafour a4paper,12pt,\else a5paper,10pt,\fi
onecolumn,oneside,article,
british%
]{memoir}
\newif\ifpublic
\publictrue 
\newcommand*{\firstdraft}{25 April 2011}

\newcommand*{\propertitle}{Affine and convex spaces\\{\large blending the analytic and geometric viewpoints}}
\newcommand*{\pdftitle}{Affine and convex spaces: blending the analytic and geometric viewpoints}
\newcommand*{\headtitle}{Affine and convex spaces}
\newcommand*{\pdfauthor}{P.G.L.  Porta Mana}
\newcommand*{\headauthor}{\ifpublic Porta Mana%
\else\autanet\ Luca\fi}
\newcommand*{\reporthead}{Tech. rep. Perimeter Institute pi-other-212}
\usepackage[T1]{fontenc} 
\input{glyphtounicode} \pdfgentounicode=1
\usepackage[utf8]{inputenx}

\usepackage{textcomp}
\usepackage[normalem]{ulem}
\usepackage{amsmath}
\usepackage{mathtools}
\usepackage{empheq}

\usepackage{framed}
\usepackage{amssymb}
\usepackage{amsxtra}

\usepackage[main=british,french,italian,german,swedish,latin,esperanto]{babel}

\newcommand*{\langlatin}{\foreignlanguage{latin}}

\usepackage[autostyle=false,autopunct=false,english=british]{csquotes}
\setquotestyle{american}

\usepackage{amsthm}

\theoremstyle{remark}

\newtheoremstyle{innote}{\parsep}{\parsep}{\footnotesize}{}{}{}{0pt}{}
\theoremstyle{innote}

\usepackage[shortlabels,inline]{enumitem}
\SetEnumitemKey{para}{itemindent=\parindent,leftmargin=0pt,listparindent=\parindent,parsep=0pt,itemsep=\topsep}
\setlist[enumerate,2]{label=\alph*.}
\setlist[enumerate]{leftmargin=\parindent}
\setlist[itemize]{leftmargin=\parindent}
\setlist[description]{leftmargin=\parindent}

\usepackage[babel,theoremfont]{newpxtext}
\usepackage[bigdelims,nosymbolsc
]{newpxmath}
\useosf
\makeatletter
\def\re@DeclareMathSymbol#1#2#3#4{%
    \let#1=\undefined
    \DeclareMathSymbol{#1}{#2}{#3}{#4}}
\re@DeclareMathSymbol{\bigoplusop}{\mathop}{largesymbols}{"4C}
\re@DeclareMathSymbol{\bigotimesop}{\mathop}{largesymbols}{"4E}
\re@DeclareMathSymbol{\sumop}{\mathop}{largesymbols}{"50}
\re@DeclareMathSymbol{\prodop}{\mathop}{largesymbols}{"51}
\re@DeclareMathSymbol{\bigcupop}{\mathop}{largesymbols}{"53}
\re@DeclareMathSymbol{\bigcapop}{\mathop}{largesymbols}{"54}
\re@DeclareMathSymbol{\bigwedgeop}{\mathop}{largesymbols}{"56}
\re@DeclareMathSymbol{\bigveeop}{\mathop}{largesymbols}{"57}
\re@DeclareMathSymbol{\bigtimesop}{\mathop}{largesymbolsPXA}{"10}
\makeatother

\DeclareFontFamily{U}{egreek}{\skewchar\font'177}%
\DeclareFontShape{U}{egreek}{m}{n}{<-6>s*[1]eurm5 <6-8>s*[1]eurm7 <8->s*[1]eurm10}{}%
\DeclareFontShape{U}{egreek}{m}{it}{<->s*[1]eurmo10}{}%
\DeclareFontShape{U}{egreek}{b}{n}{<-6>s*[1]eurb5 <6-8>s*[1]eurb7 <8->s*[1]eurb10}{}%
\DeclareFontShape{U}{egreek}{b}{it}{<->s*[1]eurbo10}{}%
\DeclareSymbolFont{egreeki}{U}{egreek}{m}{it}%
\SetSymbolFont{egreeki}{bold}{U}{egreek}{b}{it}
\DeclareSymbolFont{egreekr}{U}{egreek}{m}{n}%
\SetSymbolFont{egreekr}{bold}{U}{egreek}{b}{n}
\DeclareFontFamily{U}{egreekx}{\skewchar\font'177}
\DeclareFontShape{U}{egreekx}{m}{n}{%
       <-7.5>s*[0.9]euex7%
    <7.5-8.5>s*[0.9]euex8%
    <8.5-9.5>s*[0.9]euex9%
    <9.5->s*[0.9]euex10%
}{}
\DeclareSymbolFont{egreekx}{U}{egreekx}{m}{n}
\DeclareMathSymbol{\sumop}{\mathop}{egreekx}{"50}
\DeclareMathSymbol{\prodop}{\mathop}{egreekx}{"51}
\DeclareMathSymbol{\coprodop}{\mathop}{egreekx}{"60}
\makeatletter
\def\sum{\DOTSI\sumop\slimits@}
\def\prod{\DOTSI\prodop\slimits@}
\def\coprod{\DOTSI\coprodop\slimits@}
\makeatother
 \DeclareMathSymbol{\partialup}{\mathalpha}{egreekr}{"40}
 \DeclareMathSymbol{\epsilon}{\mathalpha}{egreeki}{"0F}
 \DeclareMathSymbol{\kappa}{\mathalpha}{egreeki}{"14}
 \DeclareMathSymbol{\lambda}{\mathalpha}{egreeki}{"15}
 \DeclareMathSymbol{\mu}{\mathalpha}{egreeki}{"16}
 \DeclareMathSymbol{\nu}{\mathalpha}{egreeki}{"17}
 \DeclareMathSymbol{\psi}{\mathalpha}{egreeki}{"20}
 \let\varkappa\kappa
 \DeclareMathSymbol{\varDelta}{\mathalpha}{egreeki}{"01}
 \DeclareMathSymbol{\varEpsilon}{\mathalpha}{egreeki}{"45}
 \DeclareMathSymbol{\varIota}{\mathalpha}{egreeki}{"49}
 \DeclareMathSymbol{\deltaup}{\mathalpha}{egreekr}{"0E}
  \DeclareMathSymbol{\piup}{\mathalpha}{egreekr}{"19}

\renewcommand\sfdefault{uop}
\DeclareMathAlphabet{\mathsf}  {T1}{\sfdefault}{m}{sl}
\SetMathAlphabet{\mathsf}{bold}{T1}{\sfdefault}{b}{sl}

\usepackage[scaled=0.84]{DejaVuSansMono}

\usepackage{mathdots}

\usepackage[usenames]{xcolor}
\definecolor{mybluishpurple}{RGB}{51,34,136}
\definecolor{myblue}{RGB}{136,204,238}
\definecolor{mybluishgreen}{RGB}{68,170,153}
\definecolor{mygreen}{RGB}{17,119,51}
\definecolor{mygreenishyellow}{RGB}{153,153,51}
\definecolor{myyellow}{RGB}{221,204,119}
\definecolor{myred}{RGB}{204,102,119}
\definecolor{mypurplishred}{RGB}{136,34,85}
\definecolor{myreddishpurple}{RGB}{170,68,153}
\definecolor{mygrey}{RGB}{221,221,221}
\colorlet{shadecolor}{mygrey}

\usepackage{bm}
\usepackage{microtype}

\usepackage[backend=biber,mcite,
citestyle=authoryear-comp,bibstyle=pglpm-authoryear,autopunct=false,sorting=ny,sortcites=false,natbib=false,maxcitenames=1,maxbibnames=8,minbibnames=8,giveninits=true,uniquename=false,uniquelist=false,maxalphanames=1,block=space,hyperref=true,defernumbers=false,useprefix=true,sortupper=false,language=british,parentracker=false]{biblatex}
\DeclareSortingScheme{ny}{\sort{\field{sortname}\field{author}\field{editor}}\sort{\field{year}}}
\DefineBibliographyExtras{british}{\def\finalandcomma{\addcomma}}
\renewcommand*{\finalnamedelim}{\addcomma\space}
\DeclareDelimFormat{multicitedelim}{\addsemicolon\space}
\DeclareDelimFormat{compcitedelim}{\addsemicolon\space}
\DeclareDelimFormat{postnotedelim}{\space}
\renewcommand{\bibfont}{\footnotesize}
\defbibheading{bibliography}[\bibname]{\section*{#1}\addcontentsline{toc}{section}{#1}
}
\newcommand*{\citep}{\parencites}
\newcommand*{\citey}{\parencites*}
\renewcommand*{\cites}{\parencites}
\providecommand{\href}[2]{#2}

\newcommand*{\amp}{\&}

\newcommand*{\subtitleproc}[1]{}

\usepackage{graphicx}
\usepackage{wrapfig}

\usepackage[hypertexnames=false]{hyperref}
\usepackage[depth=4]{bookmark}
\hypersetup{colorlinks=true,bookmarksnumbered,pdfborder={0 0 0.25},citebordercolor={0.2 0.1333 0.5333},
citecolor=mybluishpurple,linkbordercolor={0.0667 0.4667 0.2},
linkcolor=mypurplishred,urlbordercolor={0.5333 0.1333 0.3333},
urlcolor=mygreen,breaklinks=true,pdftitle={\pdftitle},pdfauthor={\pdfauthor}}
\usepackage[vertfit=local]{breakurl}

\ifafour\setstocksize{297mm}{210mm}
\else\setstocksize{210mm}{5.5in}
\fi
\settrimmedsize{\stockheight}{\stockwidth}{*}
\setlxvchars[\normalfont] 
\setxlvchars[\normalfont]
\ifafour\settypeblocksize{*}{32pc}{1.618} 
\else\settypeblocksize{*}{26pc}{1.618}
\fi
\setulmargins{*}{*}{1}
\setlrmargins{*}{*}{*}
\setheadfoot{\onelineskip}{2.5\onelineskip}
\setheaderspaces{*}{2\onelineskip}{*}
\setmarginnotes{2ex}{10mm}{0pt}
\checkandfixthelayout[nearest]
\fixpdflayout
\pdfmapline{+dummy-space <dummy-space.pfb}\pdfinterwordspaceon

\newcommand*{\asudedication}[1]{%
{\par\centering\textit{#1}\par}}
\newenvironment{acknowledgements}{\section*{Thanks}\addcontentsline{toc}{section}{Thanks}}{\par}
\makeatletter\renewcommand{\appendix}{\par
  \bigskip{\centering
   \interlinepenalty \@M
   \normalfont
   \printchaptertitle{\sffamily\appendixpagename}\par}
  \setcounter{section}{0}%
  \gdef\@chapapp{\appendixname}%
  \gdef\thesection{\@Alph\c@section}%
  \anappendixtrue}\makeatother
\counterwithout{section}{chapter}
\setsecnumformat{\upshape\csname the#1\endcsname\quad}
\setsecheadstyle{\large\bfseries\sffamily%
\centering}
\setsubsecheadstyle{\bfseries\sffamily%
\raggedright}
\setsubsecindent{0pt}
\setbeforeparaskip{-1.5ex plus 1ex minus .2ex}
\setparaheadstyle{\bfseries\sffamily%
\raggedright}
\newcommand{\addchap}[1]{\chapter*[#1]{#1}\addcontentsline{toc}{chapter}{#1}}
\newcommand{\addsec}[1]{\section*{#1}\addcontentsline{toc}{section}{#1}}

\copypagestyle{manaart}{plain}
\makeheadrule{manaart}{\headwidth}{0.5\normalrulethickness}
\makeoddhead{manaart}{%
{\footnotesize
\scshape\headauthor}}{}{{\footnotesize\sffamily%
\headtitle}}
\makeoddfoot{manaart}{}{\thepage}{}
\newcommand*\autanet{\includegraphics[height=\heightof{M}]{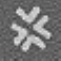}}
\definecolor{mygray}{gray}{0.333}
\iftypodisclaim%
\ifafour\newcommand\addprintnote{\begin{picture}(0,0)%
\put(245,149){\makebox(0,0){\rotatebox{90}{\tiny\color{mygray}\textsf{This
            document is designed for screen reading and
            two-up printing on A4 or Letter paper}}}}%
\end{picture}}
\else\newcommand\addprintnote{\begin{picture}(0,0)%
\put(176,112){\makebox(0,0){\rotatebox{90}{\tiny\color{mygray}\textsf{This
            document is designed for screen reading and
            two-up printing on A4 or Letter paper}}}}%
\end{picture}}\fi
\makeoddfoot{plain}{}{\makebox[0pt]{\thepage}\addprintnote}{}
\else
\makeoddfoot{plain}{}{\makebox[0pt]{\thepage}}{}
\fi
\makeoddhead{plain}{}{}{\footnotesize\reporthead}

\pretitle{\begin{center}\LARGE\sffamily%
\bfseries}
\posttitle{\bigskip\end{center}}

\makeatletter\newcommand*{\atf}{\includegraphics[
totalheight=\heightof{@}]{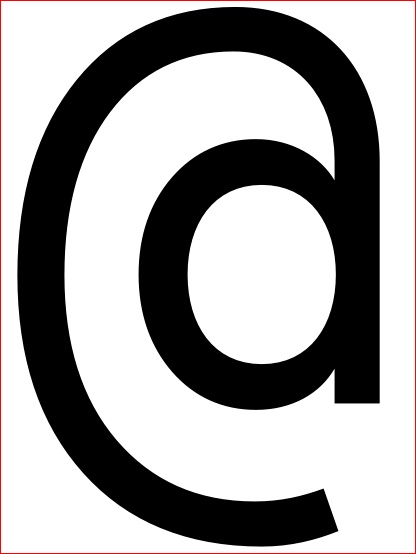}}\makeatother

\providecommand{\epost}[1]{\texttt{\footnotesize\textless#1\textgreater}}
\providecommand{\email}[2]{\href{mailto:#1ZZ@#2 ((remove ZZ))}{#1\protect\atf#2}}

\preauthor{\vspace{-0.5\baselineskip}\begin{center}
\normalsize\sffamily%
\lineskip  0.5em}
\postauthor{\par\end{center}}
\predate{\DTMsetdatestyle{mydate}\begin{center}\footnotesize}
\postdate{\end{center}\vspace{-\medskipamount}}
\usepackage[british]{datetime2}
\DTMnewdatestyle{mydate}%
{
}
\DTMsetdatestyle{mydate}

\setfloatadjustment{figure}{\footnotesize}
\captiondelim{\quad}
\captionnamefont{\footnotesize\sffamily%
}
\captiontitlefont{\footnotesize}
\firmlists*
\midsloppy

\title{\propertitle
}
\author{\ifpublic%
P.G.L. Porta\,Mana%
\else Luca\fi
\quad
\epost{\email{pgl}{portamana.org}}%
}

\date{\firstdraft; updated \today}

\newcommand*{\delt}{\deltaup}
\newcommand*{\RR}{\bm{\mathrm{R}}}
\newcommand*{\defd}{\coloneqq}

\newcommand*{\mimplies}{\mathrel{\;\Longrightarrow\;}}
\newcommand*{\inn}{\cdot}
\newcommand*{\comp}{\circ}
\providecommand{\varparallel}{\ensuremath{\mathbin{/\mkern-3mu/}}}
\renewcommand{\le}{\leqslant}
\DeclarePairedDelimiter\clcl{[}{]}

\DeclarePairedDelimiter\opop{]}{[}
\DeclarePairedDelimiter\abs{\lvert}{\rvert}
\DeclarePairedDelimiter\set{\{}{\}}
\newcommand*{\p}{\mathrm{P}}

\newcommand*{\sect}{\S}
\newcommand*{\sects}{\S\S}
\newcommand*{\chap}{ch.}%
\newcommand*{\chaps}{chs}%
\newcommand*{\fig}{fig.}%

\newcommand*{\etc}{{etc.}}
\newcommand*{\ie}{{i.e.}}
\newcommand*{\eg}{{e.g.}}

\newcommand*{\viz}{{viz}}

\newcommand*{\bd}{\hspace{0pt}}%

\newcommand*{\tsum}{\mathop{\textstyle\sum}\nolimits}

\definecolor{notecolour}{RGB}{68,170,153}

\newcommand*{\ya}{\lambda}
\newcommand*{\yb}{\mu}
\newcommand*{\yotimes}{\,}
\newcommand*{\yoplu}{\boxplus}
\newcommand*{\yoplus}{\dotplus}

\newcommand*{\yu}{\bm{u}}
\newcommand*{\yv}{\bm{v}}
\newcommand*{\ySS}{\mathcal{S}}

\newcommand*{\yO}[1]{\mathcal{P}(#1)}

\DeclareMathOperator{\aff}{aff}
\newcommand*{\afs}{A}

\newcommand*{\ydu}{d_{\mathrm{u}}}
\newcommand*{\yom}{v}
\newcommand*{\yoz}{\yom_0}
\newcommand*{\you}{\yom_1}
\newcommand*{\ymP}{m_\text{P}}
\newcommand*{\ymC}{m_\text{C}}
\newcommand*{\yF}[1]{F_{#1}}
\newcommand*{\st}{\mid}

\firmlists
\begin{document}
\captiondelim{\quad}\captionnamefont{\footnotesize}\captiontitlefont{\footnotesize}
\selectlanguage{british}\frenchspacing

\maketitle
\ifpublic
\abstractrunin
\abslabeldelim{}
\renewcommand*{\abstractname}{}
\setlength{\absleftindent}{0pt}
\setlength{\absrightindent}{0pt}
\setlength{\abstitleskip}{-\absparindent}
\begin{abstract}\labelsep 0pt%
  \noindent This is a short introduction to affine and convex spaces,
  written especially for physics students. It summarizes different
  elementary presentations available in the mathematical literature, and
  blends analytic- and geometric-flavoured presentations. References are
  also provided, as well as a brief discussion of Grassmann spaces and an
  example showing the relevance and usefulness of affine spaces in
  Newtonian physics.
\par
\noindent
{\footnotesize \textsc{pacs}: 02.40.Dr,02.40.Ft,45.20.-d}\qquad%
{\footnotesize \textsc{msc}: 14R99,51N10,52A20}%
\end{abstract}\fi

\selectlanguage{british}\frenchspacing
\asudedication{\small to Louise}


\section{Spaces that deserve more space}
\label{sec:genintro}

Scientists and science students of different fields are very familiar, in
various degrees of sophistication, with vector spaces. Vectors are used to
model places, velocities, forces, generators of rotations, electric fields,
and even quantum states. Vector spaces are among the building blocks of
classical mechanics, electromagnetism, general relativity, quantum theory;
they constitute therefore an essential part of physics and mathematics
teaching.

Physicists also like to repeat, with Newton \parentext{\cite*[Liber~III,
re\-gu\-la~I]{newton1687_r1726}; \cite[\sect~293]{truesdelletal1960}}, a maxim
variously attributed \citep{thorburn1918} to Ockham or Scotus:
\enquote{\emph{\langlatin{frustra fit per plura quod fieri potest per
      pauciora}}}. Applied to the connexion between mathematics and
physics, it says that we should not model a physical phenomenon by means of
a mathematical object that has more structure than the phenomenon itself.
But this maxim is forsaken in the case of vector spaces, for they are
sometimes used where other spaces, having less structure, would suffice. A
modern example is given by quantum theory, where \enquote{pure states} are usually
represented as (complex) vectors; but the vectors $\bm{\psi}$ and
$\lambda\bm{\psi}$, $\lambda \ne 0$, represent the same state, and the null
vector represents none. Clearly the vector-\bd space structure is redundant
here. In fact, pure quantum states should more precisely be seen as points
in a complex projective space
\citep[\sect~1.3.1]{haag1992_r1996}{bengtssonetal2006_r2017}.


Another example is the notion of reference frame in classical
gal\-i\-le\-ian-\bd relativistic mechanics: such a frame is often modelled
as a vector space, wherein we describe the \emph{place} occupied by a small
body by its \enquote{position vector} with respect to some origin. But suppose
that I choose two places, for example, on a solar-\bd system scale, those occupied
by Pluto and Charon at a given time; and I ask you: what is the \emph{sum}
of these places? This question does not make very much sense; and even if
you associate two vectors to the two places and then perform a formal sum
of those vectors, the resulting place is devoid of any physical meaning.
Thus, even if we usually model places as vectors, it is clear that the
mathematical structure given by vector addition has no physical counterpart
in this case.

On the other hand, I can ask you to determine a place in between the places
occupied by Pluto and Charon such that its distances from the two planets
are in an inverse ratio as the planets' masses $\ymP$, $\ymC$; in other
words, their mass centre. You can obtain this place unambiguously, and it
also has a physical meaning: it moves as the place occupied by a body with
mass $\ymP+\ymC$ under the total action of the forces acting on Pluto and
Charon. It turns out that the operation of assigning a mass-\bd centre does
not really need the concept of distance, and can be modelled in a space
that has less structure, and is therefore more general, than a vector
space: an \emph{affine space}.

Affine spaces have geometrically intuitive properties and are not more
difficult to understand than vector spaces. But they are rarely taught to
physics students; and when they are, they are presented as by-products of vector
spaces. This is reflected in textbooks of mathematical methods in physics.
Amongst the old and new, widely and less widely known textbooks that I
checked
\citep{courant1924_r1966,jeffreysetal1946_r1950,schouten1951_r1989,fluegge1955,fluegge1956,wilf1962_r1978,arfkenetal1966_r2005,boas1966_r1983,reedetal1972_r1980,choquetbruhatetal1977_r1996,marsdenetal1983_r2007,geroch1985,bambergetal1988_r1990,rileyetal1998_r2002,hassani1999_r2009,szekeres2004},
only Bamberg \amp\ Sternberg \citey{bambergetal1988_r1990}, Szekeres
\citey{szekeres2004}, and obviously Schouten \citey{schouten1951_r1989}
give appropriate space to affine spaces; almost all others do not even
mention affine spaces at all, although all of them obviously present the
theory of vector spaces.

The students who have heard about affine spaces and would like to know more
about them will find heterogeneous material, scattered for the most part in
books and textbooks about general geometry. Part of this material has an
analytic flavour, part a geometrical flavour; and to get a more all-round
view some patch-work is needed. It is the purpose of these notes to offer
such patch-work, emphasizing the dialogue between the analytic and the
synthetic-\bd geometric presentations, and to offer some references. An
intuitive knowledge of basic geometrical notions is assumed.

Closely related to affine spaces are \emph{convex spaces}. These are also
geometrically very intuitive, and are ubiquitous in convex analysis and
optimization. Although their range of application in physics is maybe
narrower than that of affine and vector spaces, they appear naturally in
probability theory and therefore in statistical physics, be it the
statistical mechanics of mass-points, fields, or continua; and they are of
utmost importance in quantum theory, being behind many of its non-\bd
classical properties. Quantum theory is indeed only a particular example of
a general plausibilistic physical theory, a particular case of a
statistical model; and convex spaces are the most apt spaces to study the
latter.

Students who have heard about and are interested in the general theory of
convex spaces will find even less, and more hidden, material than for
affine spaces. These notes offer some references and a general overview of
convex spaces, too.

At the end of these notes I shall briefly discuss \emph{Grassmann spaces},
which generalize affine spaces in many remarkable ways, and give an example
of application of affine spaces in Newtonian mechanics, related to the
previous discussion about Pluto and Charon. Extended application examples
for convex spaces are left to a future note.

\section{Affine spaces}
\label{sec:affintro}

\subsection{Analytic point of view}
\label{sec:analyt_view}

\paragraph{Affine combinations}
An \emph{affine space} is a set of \emph{points} that is closed
under an operation, \emph{affine combination}, mapping pairs of points
$(a,b)$ and pairs of real numbers $(\ya,\yb)$ summing up to one to
another point $c$ of the space:
\begin{equation}
  \label{eq:affine_sum}
(a,b,\ya,\yb)\mapsto c=\ya \yotimes a \yoplu \yb \yotimes b,
\quad \ya,\yb \in \RR,\quad \ya+\yb=1.
\end{equation}
The intuitive properties of this operation, including the extension to more
than two points, I do not list here. It behaves in a way similar to scalar
multiplication followed by vector addition in a vector space; but as the
symbol 
\enquote{$\yoplu$} in place of \enquote{$+$} reminds us,
\enquote{multiplication} of a point by a number and \enquote{sum} of two
points are undefined operations in an affine space: only the combination
above makes sense. This operation has a geometric meaning which will be
explained in \sect~\ref{sec:geom_view}. One usually writes simply
$\ya a + \yb b$, a notation that we shall follow. Whereas a vector space
has a special vector: the null vector, an affine space has no special
points and is therefore more general than a vector space.

\paragraph{Affine basis}

A set of points is \emph{affinely independent} if none of them can be
written as an affine combination of the others. The maximum number of
affinely independent points minus one defines the dimension of the affine
space. An \emph{affine basis} is a maximal set of affinely independent
points. Any point of the space can be uniquely written as an affine
combination of basis points, and the coefficients can be called the
\emph{weights} of the point with respect to that basis. A choice of basis
allows us to baptize each point with a numeric name made of $n$ reals
summing up to one, where $n$ is the dimension of the space plus one. This
$n$-tuple can be represented by a column
matrix
. An affine combination of two or more points corresponds to a sum of their
matrices multiplied by their respective coefficients. For particular affine
spaces whose points are already numbers, like the real line $\RR$, such
baptizing ceremonies are usually superfluous.

\paragraph{Affine subspaces}

Some subsets of an affine space are affine spaces themselves, of lower
dimensionality. Given two points $a_1$, $a_2$, the \emph{line} $a_1a_2$
through them is the locus of all points obtained by their affine
combinations for all choices of coefficients $(\ya_1, \ya_2)$, $\ya_1
+\ya_2 =1$. Given three affinely independent points $a_1$, $a_2$, $a_3$,
the \emph{plane} $a_1a_2a_3$ through them is the locus of all points
obtained by their affine combinations for all choices of coefficients
$(\ya_1, \ya_2, \ya_3)$, $\tsum_i \ya_i=1$. And so on for $n+1$ points and
$n$-dimensional planes, the latter called $n$-planes for short. All these
are affine subspaces: a line, of dimension one; a plane, of dimension two;
\etc{} In general, given a finite set of points $\set{a_1,\dotsc,a_r}$, not
necessarily affinely independent, their \emph{affine span}
$\aff\set{a_1,\dotsc,a_r}$ is the smallest affine subspace containing
them. It is simply the locus of all points obtained by affine combinations
of the $\set{a_i}$ for all possible choices of coefficients.

Given four points $a_1$, $a_2$, $b_1$, $b_2$, the lines $a_1a_2$ and
$b_1b_2$ are said to be \emph{parallel}, written $ a_1a_2 \varparallel
b_1b_2$, according to the following definition:
\begin{equation}
  \label{eq:paral_line}
  a_1a_2 \varparallel b_1b_2 \iff b_2=
b_1-\ya a_1 +\ya a_2 \text{ for some $\ya$}.
\end{equation}
For two planes $a_1a_2a_3$ and $b_1b_2b_3$ to be parallel we must have
\begin{multline}
  \label{eq:paral_plane}
  a_1a_2a_3 \varparallel b_1b_2b_3 \iff{}\\
\left\{
  \begin{aligned}
    b_2&= b_1 -(\ya+\yb) a_1+ \ya a_2 +\yb a_3 \quad\text{for some $\ya$, $\yb$},
\\
b_3&= b_1 -(\ya'+\yb') a_1+\ya' a_2 +\yb' a_3\quad\text{for some $\ya'$, $\yb'$}.
  \end{aligned}
\right.
\end{multline}
And so on. We shall see later that these notions coincide with the usual
geometric ones. Geometrically, affine dependence means collinearity,
coplanarity, \etc

\paragraph{Affine mappings}
\label{sec:affmaps}

An \emph{affine mapping} or \emph{affinity} from one affine space to
another or to itself is a mapping $F$ that preserves affine combinations:
\begin{equation}
  \label{eq:affine_map}
  F(\tsum_i \ya_i a_i) = \tsum_i \ya_i F(a_i),
\qquad \tsum_i \ya_i = 1;
\end{equation}
it therefore maps $r$-planes into $t$-planes, where $t \le r$, and mutually
parallel objects into mutually parallel objects. In the following we shall
often use the summation convention and omit the normalization condition
when clear from the context.

Introducing two affine bases in the domain and range of an affine mapping
it is easy to see that it can be represented by a stochastic
$(m+1,n+1)$-matrix (\ie, with columns summing up to one), operating on a
point through multiplication by the latter's column matrix; $n$ and $m$ are
the dimensions of domain and range. This representation is basis-dependent.
The rank of the matrix, which is basis-independent (and obviously smaller
than $n+1$ and $m+1$), is equal to the dimension of the image of the domain
plus one. When domain, range, and the image of the domain have the same
dimension the matrix is square and its non-vanishing determinant, also
basis-independent, is the \emph{ratio} of the hypervolumes determined by
the image of the first space's basis and that formed by the second space's
basis; more on ratio of hypervolumes in \sect~\ref{sec:geom_view}.

We can define affine combinations of the affinities between two affine
spaces in a canonical way: $(\ya F + \yb G)(a)\defd \ya F(a) +\yb G(a)$ for
any two affinities $F$, $G$ with same $n$-dimensional domain and
$m$-dimensional range (note how the expression \enquote{$\ya F(a)$} by itself has no meaning). The set of these mappings is therefore an affine space
itself, of dimension $m(n+1)$.

\paragraph{Affine forms}

An affinity from an $n$-dimensional affine space to the real line $\RR$ can
be represented by a single-row matrix with $n$ entries, instead of a
$(2,n)$ matrix, because as already said the points of the reals can
numerically represent themselves without the need of an affine
basis
. The matrix representation is still dependent on a choice of basis in the
domain affine space, though. Such affinities are called \emph{affine forms}
or simply \emph{forms}. Their set is an affine space; in fact, it is even a
vector space owing to the vector structure of the reals; its dimension in
both cases is $n+1$, thus larger than that of the original affine space (for this reason I find the name \enquote{dual space}, used by some,
inappropriate). The action of an affine form $v$ on a point $a$ will be
denoted by $v \inn a$.

Choose a basis $(e^i)$ in an affine space. In the space of forms, seen
as a \emph{vector} space, we can then choose a vector basis $(d_j)$
such that $d_j \inn e^i = \delt_j^i$; the $d_j$ are called \emph{dual
  forms} of the basis $(e^i)$. The set $\set{d_j}$ is however
insufficient as a basis if we see the space of forms as an affine space:
it has to be augmented by another form, like the null-form $d_0\colon a
\mapsto 0$ or the unit-form $\ydu\colon a \mapsto 1$. Here we choose the
former; $(d_0, d_j)$ is thus an affine basis in the space of forms.

\begin{figure}[!b]
  \centering
\includegraphics[width=0.48\columnwidth]{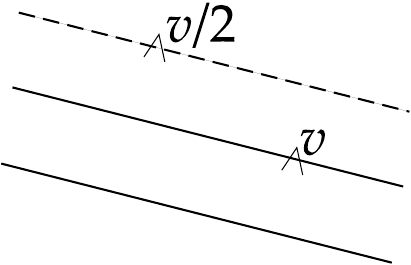}\hfill
\includegraphics[width=0.48\columnwidth]{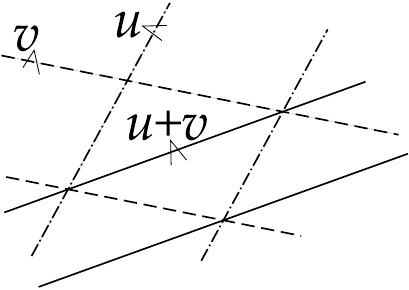}
\caption{Scalar multiplication and addition of affine forms}
\label{fig:forms_op}
\end{figure}

A non-constant affine form can be geometrically seen as a family of
parallel hyperplanes in the affine space: the form has a constant value on
each hyperplane. This family is usually iconized by drawing only two
hyperplanes: one, unmarked, where the form has value zero, and one, marked
by \eg\ a tick, where it has value one. A constant form has no such
hyperplanes of course. The $r$-multiple of a form has its unity hyperplane
at a distance $1/r$ times the original distance from the zero hyperplane.
If you wonder what I mean with \enquote{distance}, given that no such notion is
defined in an affine space, please read the next section. The sum of two
forms is a form whose unity hyperplane passes through the intersections of
their zero and unity hyperplanes, and whose parallel zero hyperplane passes
through the intersection of their zero hyperplanes. See
fig.~\ref{fig:forms_op} and the nice illustrations in Burke
\citey{burke1985_r1987,burke1995}.

\subsection{Geometric point of view}
\label{sec:geom_view}

\paragraph{Parallelism and translations}

From a geometric point of view, an affine space is based on the notions
of point, line, plane, space, hyperplane, an so on, and the notion of
(Euclidean) parallelism. I shall take these notions, that can be
axiomatized in many different ways, for granted. Note that the notions of
distance and angle are undefined.

\begin{figure}[!b]
  \centering
\includegraphics[width=0.65\columnwidth]{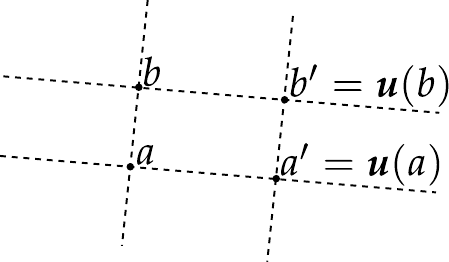}
\caption{Construction of the image $b'=\yu(b)$ of $b$ by the translation
  determined by $a$ and $a'=\yu(a)$}
\label{fig:paral_const}
\end{figure}

Affine mappings between affine spaces are those that preserve the relation of
parallelism: they map pairs of parallel objects, like lines or hyperplanes,
into pairs of parallel objects. A special group of affine transformations
of an affine space into itself are those that map every object into another
parallel to, and of the same dimension as, the original one. They are
called \emph{translations}. To specify a translation $\yu$ we only need to
assign a point $a$ and its image $a'\defd\yu(a)$. The image $b'\defd\yu(b)$
of any other point $b$ outside of the line $aa'$ is determined by requiring
that the line $bb'$ be parallel to $aa'$ and the line $a'b'$ to $ab$, as in
fig.~\ref{fig:paral_const}. The segment $bb'$ can then be used to construct
the image of other points on the line $aa'$; from this construction it is
clear that the case of a one-\bd dimensional affine space requires a
different approach. The translations form a commutative group, the
identity being the null translation $\bm{0}\colon a \mapsto a$, and the
inverse of $\yu$ being the translation $-\yu$ determined by $\yu(a)$ and
its image $a=-\yu[\yu(a)]$. The action of this group on the affine space is
transitive, faithful, and free.

\begin{figure}[!b]
  \centering
\includegraphics[width=\columnwidth]{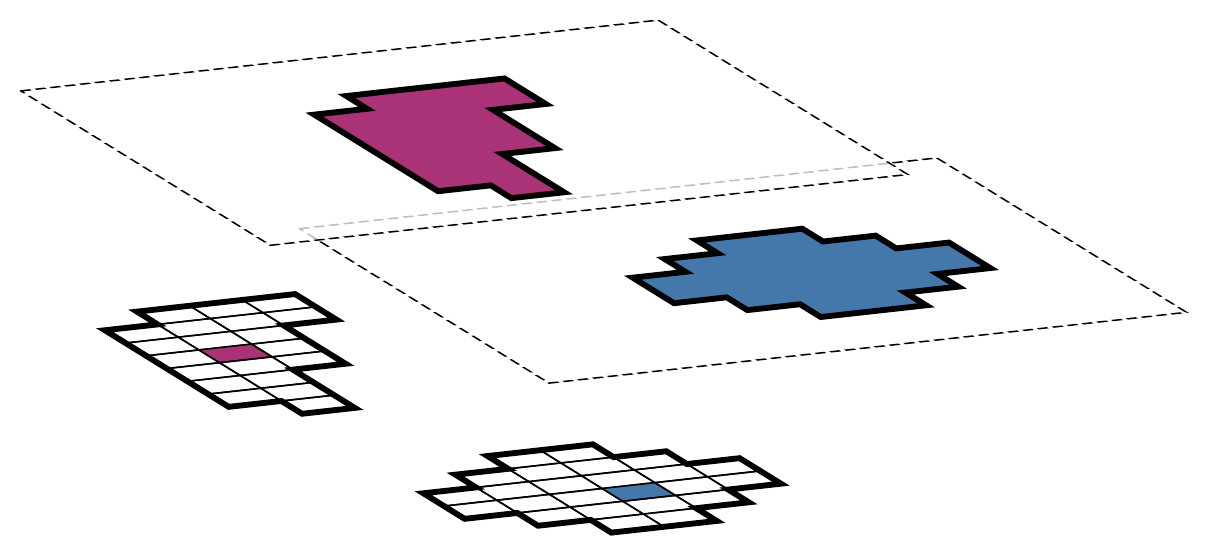}
\caption{The orange and blue areas on the two parallel planes are in the
  ratio $21/24$, as the decomposition into smaller equal rectangles shows
  underneath. With a limit construction we can compare parallel areas with
  curvilinear boundaries}
\label{fig:ratio_areas}
\end{figure}

\paragraph{Ratio of $n$-areas}

If the point $a_1$ is the image of $a$ under $\yu$, and $a_2$ the image of
$a$ under a double application,
$\yu'\defd 2\yu \defd \yu+\yu \defd \yu\comp\yu$, of the same translation,
we can say that the oriented segment $\overrightarrow{aa_2}$ is twice
$\overrightarrow{aa_1}$, or that the latter is half the former, and we can
write $\yu=\yu'/2$. Generalizing this construction we can define rational
multiples of a translation, and thence generic real multiples $\ya\yu$,
$\ya\in\RR$, through a Dedekind-\bd section-\bd like construction
\citep[\sect~13.3]{coxeter1961_r1969}. Negative values indicate a change in
orientation. Translations form therefore a vector space over the reals,
sometimes called the \emph{translation space} of the original affine space,
and they allow us to speak of the \emph{ratio} of two lengths along
parallel lines (but not along non-parallel ones), of two areas on parallel
planes, and so on with $n$-areas, up to the ratio of any two hypervolumes.
The procedure is to divide the first $n$-area into smaller and smaller
equal $n$-rectangles, and to see how many of them are in the limit needed
to fill, by translation, the second $n$-area; see the example in
fig.~\ref{fig:ratio_areas}. The ratio between two hypervolumes provides a
geometric definition of the determinant of an affinity, defined in
\sect~\ref{sec:affmaps} in terms of the matrix representing the affinity.
It should now be clear what we meant, in the section about affine forms,
when we said that the distance between two parallel hyperplanes is $r$
times the distance between two other parallel hyperplanes: draw an
arbitrary line intersecting all these hyperplanes; then the segment
intercepted on the line by the first two hyperplanes and that intercepted
by the last two hyperplanes are in the ratio $r$.

\subsection{Relation between analytic and geometric points of view}
\label{sec:relation_an_geom}

The action of a translation $\yu$ on the point $b$ is usually denoted by
$b + \yu \defd \yu(b)= c'$. We also write $\yu= c-b$ to denote the fact
that $\yu$ is uniquely determined by some point $b$ and its image $c$.
Then, by what we said in \sect~\ref{sec:geom_view}, the translation
$\ya\yu= \ya(c-b)$ maps $b$ to a point $c'$ such that
$\overrightarrow{bc'}$ is $\ya$ times $\overrightarrow{bc}$ (negative
values indicating a change in orientation). The action of the same
translation $\ya\yu$ on the point $a$ can then be written $a+\ya(c-b)$. Given
another translation $\yb\yv= \yb(d-b)$, the action of the composite
translation $\ya\yu+\yb\yv$ on $a$ can be written as
$a+\ya(c-b)+\yb(d-b)$
. Generalizing this we obtain expressions which are formal sums of affine
points with coefficients summing up to unity. This provides a link between
the geometric and analytic presentations of an affine space: any affine
combination $\ya_i a^i$ can be written and interpreted
as the image $a + \sum_i\ya_i (a^i -a)$ of some point $a$ under the
composition of the translations $\ya_j (a^j -a)$, and vice versa. Note
again that the expression \enquote{$a-b$} does not denote a point of the
affine space but a particular mapping (translation) onto the space.

An expression like \enquote{$a-b-c$} has no meaning in an affine space, not
even in terms of translations. In \sect~\ref{sec:generalizations}, however,
we will briefly discuss spaces for which such an expression makes sense and
moreover the difference between points and translations disappears.

\begin{figure}[!t]
  \centering
\includegraphics[width=0.65\columnwidth]{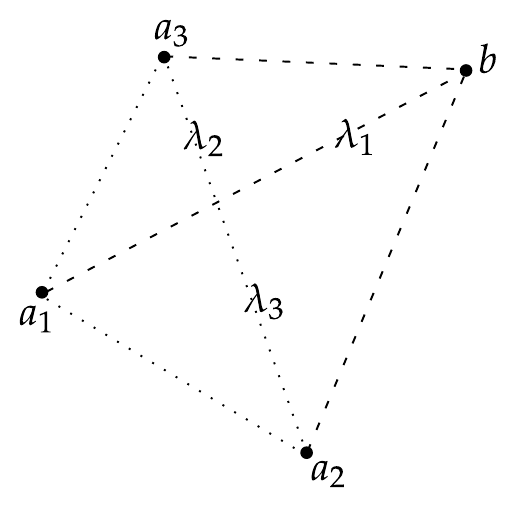}
\caption{Geometric meaning of the affine combination $b=\ya_1 a_1+\ya_2
  a_2+\ya_3 a_3$: the ratios of the triangles $a_2a_3b$, $a_1a_3b$, and
  $a_1a_2b$ to $a_1a_2a_3$ are $\abs{\ya_1}$, $\abs{\ya_2}$, and
  $\abs{\ya_3}$. The coefficient of $a_1$ is negative: $\ya_1<0$, because
  $b$ and $a_1$ lie on opposite sides of the line $a_2a_3$}
\label{fig:geom_plane}
\end{figure}
A geometric interpretation of the affine combination $b=\ya_1 a_1+\ya_2
a_2$ is that $b$ is a point on the line determined by $a_1$ and $a_2$ and
such that the unoriented segment $\overline{a_2b}$, \ie\ the one $a_1$ is
not generally an endpoint of, is $\ya_1$ times the segment
$\overline{a_1a_2}$, a negative ratio indicating that $b$ and $a_1$ lie on
opposite sides of $a_2$; and analogously for $\overline{a_1b}$ and $\ya_2$.
You can prove for yourself that the geometric interpretation of the
combination $b=\ya_1 a_1+\ya_2 a_2+\ya_3 a_3$, with the $a_i$ affinely
independent, is that $b$ is a point in the plane determined by the $a_i$
and such that the triangle $a_2 a_3 b$, \ie\ the one $a_1$ is not generally
a vertex of, is $\ya_1$ times the triangle $a_1a_2a_3$, the ratio being
negative if $b$ and $a_1$ lie on opposite sides of the line $a_2a_3$; and
analogously for the other triangles with $b$ as a vertex and the other
coefficients; see fig.~\ref{fig:geom_plane}.

Note again that lengths, areas, \etc\ belonging to non-parallel subspaces
cannot be directly compared. For that purpose one can use affine forms,
two-forms, twisted forms, \etc, which however will not be discussed in this
note. For those I refer you to the works of Burke
\citey{burke1985_r1987,burke1995}, Bossavit
\citey{bossavit1991,bossavit1994_r2002}, and also Schouten
\citey{schouten1951_r1989}.

\subsection{References}
\label{sec:ref_affine}

Excellent analytic and geometric introductions to affine spaces and mappings
can be found in \chap~13 of Coxeter \citey{coxeter1961_r1969}, \chap~II of
Artin \citey{artin1955}, \sect~I.1 of Burke \citey{burke1985_r1987}, and
also in \chaps~I--III of Schouten \citey{schouten1951_r1989} and in Boehm
\amp\ Prautzsch \citey{boehmetal2000}.

\section{Convex spaces}
\label{sec:convex_sp}

\subsection{Convex combinations and mixture spaces}
\label{sec:conv_comb}

\paragraph{Mixture spaces and convex spaces}

\begin{figure}[!t]
  \centering
  \includegraphics[width=10\columnwidth/10]{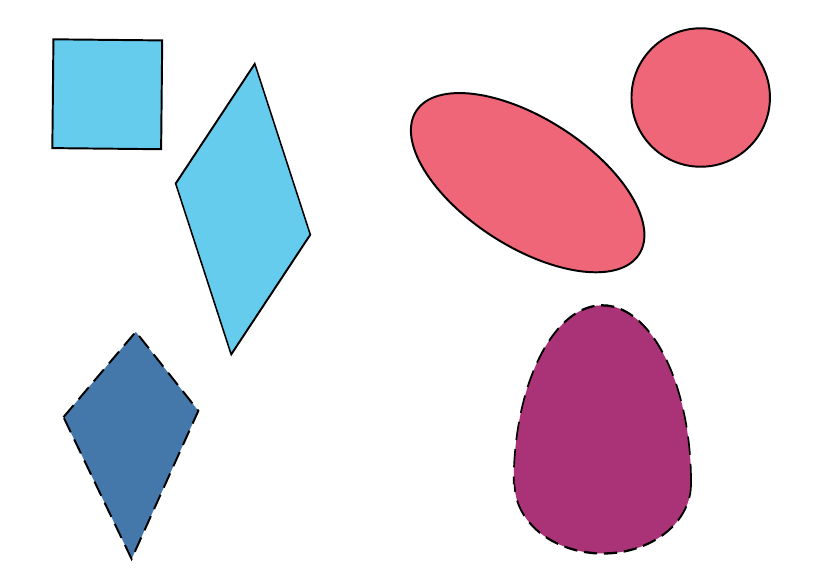}
  \caption{The two upper quadrilateral figures are the same convex space,
    whereas the lower, dashed, darker quadrilateral one is a different
    convex space; analogously for the three rounded figures}
\label{fig:convexf}
\end{figure}

A convex space is analytically defined as a set of points which is closed
under the operation of \emph{convex combination}, mapping pairs of
points and pairs of non-negative real numbers summing up to one to another
point of the space:
\begin{equation}
  \label{eq:convex_sum}
(a,b,\ya,\yb)\mapsto c=\ya \yotimes a \yoplus \yb \yotimes b,
\quad \ya,\yb \in \clcl{0,1}, \quad \ya+\yb=1.
\end{equation}
This operation satisfies additional properties, and their analysis is
interesting: Three of them,
\begin{subequations}\label{eq:mix_ax}
  \begin{gather}
  1 \yotimes a \yoplus 0 \yotimes b = a,\\
\yb \yotimes b \yoplus \ya \yotimes a =
\ya \yotimes a \yoplus \yb \yotimes b,\\
\yb \yotimes [\ya \yotimes a \yoplus (1-\ya) \yotimes b]
\yoplus (1-\yb) b =
\ya\yb \yotimes a \yoplus (1-\ya\yb) \yotimes b,
  \end{gather}
  define a \emph{mixture space}. To define a convex space, which is less
  general than a mixture space, we need two additional properties:
\begin{gather}
\label{eq:conv1}  
b \mapsto \ya\yotimes a \yoplus (1-\ya) \yotimes b
\quad\text{is injective
for all $\ya\in\opop{0,1}$ and all $a$
},
\\
\shortintertext{and}
\begin{multlined}[b][.9\columnwidth]
\label{eq:conv2}
\yb \yotimes [\ya
  \yotimes a \yoplus (1-\ya) \yotimes b] \yoplus (1-\yb) \yotimes c 
={}
\\
\shoveright{
\ya\yb
  \yotimes a \yoplus (1-\ya\yb) \yotimes \biggl[\frac{(1-\ya)\yb}{1-\ya\yb}
  \yotimes b \yoplus \frac{1-\yb}{1-\ya\yb} \yotimes c \biggr]
}
\\
\text{for all $\ya,\yb\in\clcl{0,1}$ with $\ya\yb \ne 1$.}
\end{multlined}
\end{gather}
\end{subequations}
Convex spaces are special amongst mixture spaces because they can always be
represented as convex subsets of some affine space; this property does not
need to hold for a generic mixture space
\citep{mongin2001}[\sect~VII.2]{wakker1988}. All such representations of a
convex space are isomorphic to one another, and their affine spans are also
isomorphic. This allows us to rewrite expressions like
\eqref{eq:convex_sum} as $\ya a + \yb b$ and to interpret them in the
affine sense~\eqref{eq:affine_sum}; it also allows us to speak of the
dimension of a convex space, defined as the dimension of the affine span of
any of its representations, and to speak of other notions like parallelism
and compactness. From now on we shall only consider convex rather than
mixture spaces, and finite-dimensional, compact convex spaces in
particular. See fig.~\ref{fig:convexf} for some examples of equivalent and
inequivalent convex spaces.

\paragraph{Extreme points and bases}

\begin{figure}[!t]
  \centering
  \includegraphics[width=0.9\columnwidth]{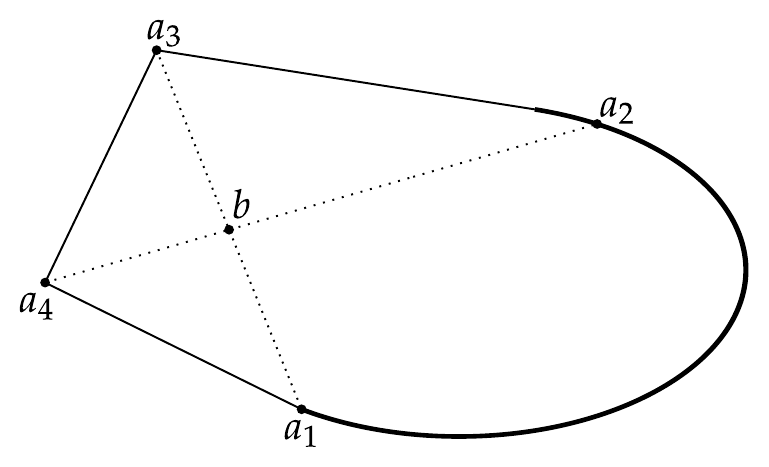}
  \caption{Example of a convex set. The points $a_1$, $a_2$, $a_3$, $a_4$
    are extreme points of the set, as well as all points on the (thicker)
    curved part of the boundary. The point $b$ can be written as a convex
    combination of extreme points in at least two different ways: as
    $a_1/2 + a_3/2$ or as $a_2/3 + 2 a_4/3$. The points $a_4$ and $a_3$ are
    faces, as is each point on the (thicker) curved part of the boundary}
\label{fig:convset}
\end{figure}

A set of points is \emph{convexly independent} if none of them can be
written as a convex combination of the others. The \emph{extreme points} of
a convex space are the convexly independent points that convexly span the
whole convex space (their set can be empty for non-compact convex
spaces). Equivalent characterizations are possible, \eg\ a point is extreme
if its exclusion from the original convex set leaves a set that is still
convex  \citep{klee1957}.
A point can generally be written as a convex combination of extreme points
in more than one way, so we cannot use them as a \enquote{convex basis} to assign
unambiguous numeric names to the other points (one can select a unique
convex combination through additional requirements, \eg\ that its weights
have maximum Shannon entropy). See fig.~\ref{fig:convset}. But through the
representation of the convex space in an affine space we can introduce an
\emph{affine basis}, whose elements can lie outside the convex space, and
write every point of the convex space uniquely as an \emph{affine}
combination of these basis elements; this affine combination will not in
general be a convex combination, \ie\ its weights can be strictly negative
or greater than unity. A weight lying in $\clcl{0,1}$ will be called
\emph{proper}, otherwise \emph{improper}.

A \emph{face} of a convex space is a subset which is also convex and which
contains all points that can be convexly combined into each of its points
\cites[\eg][\sect~XI.B]{valentine1964}[\sect~18]{rockafellar1970_r1972}; in
formulae, $F$ is a face if and only if
\begin{equation}
  \label{eq:face_purification}
\left\{  \begin{aligned}
    a_1, a_2 \in F \text{ and } a = \ya a_1 + (1-\ya)\, a_2
  &\mimplies a \in F,
    \\
  a \in F \text{ and } a = \ya a_1 + (1-\ya)\, a_2
  &\mimplies a_1, a_2 \in F.
  \end{aligned}\right.
\end{equation}
Quantum physicists call the first property (the convexity of the set)
\enquote{invariance under mixing} and the second \enquote{invariance under
  purification} \citep[\sect~1.1]{bengtssonetal2006_r2017}. See
\fig~\ref{fig:convset}. A \emph{facet} is a face of one less dimension than
the convex space.

The boundary points of a convex space can be classified according to
several other properties
\cites[\sect~XI.B]{valentine1964}[\sect~18]{rockafellar1970_r1972}[parts~IV,
XI]{valentine1964}[\sect~II.5]{alfsen1971}[\sect~5]{broendsted1983}, for
example \emph{exposed points}, which we briefly mention again in the next
section. Such properties often correspond to important physical properties
in the physical theories where convex sets find application. Examples are
thermostatics \citep{wightman1979}, where they for example indicate mixed
phases (exposed faces) and critical points (non-exposed faces), and quantum
theory
\citep{bengtssonetal2006_r2017,kimura2003,kimuraetal2004,peresetal1998}.

\begin{figure}[!pt]
  \centering
  \includegraphics[width=9.5\columnwidth/10]{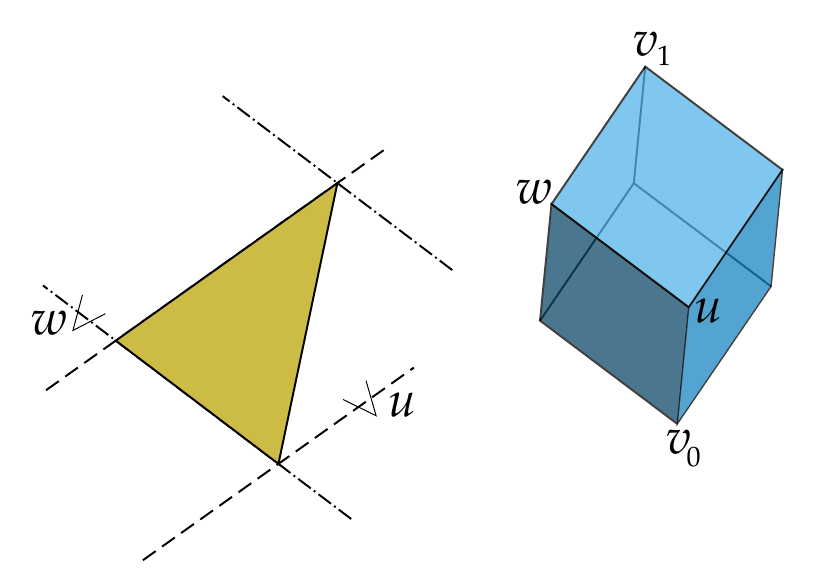}
  \includegraphics[width=9.5\columnwidth/10]{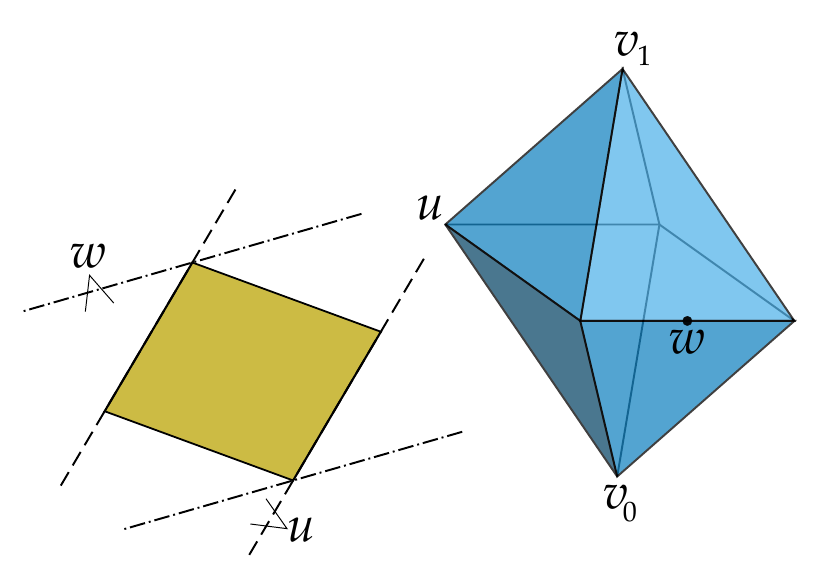}
  \caption{Two two-dimensional convex spaces, on the left, with their
    three-dimensional convex-form spaces, on the right. The convex
    forms $w$, $u$ are represented as pairs of parallel lines on the convex
  spaces and as points on the convex-form spaces. $v_0$ and $v_1$ are the
  null- and unit-forms.}
\label{fig:out_sp}
\end{figure}

A \emph{simplex} is a convex space with a number of extreme points
exceeding its dimension by one. The extreme-point decomposition of a point
of a simplex is always unique, hence a simplex' extreme points constitute a
canonical affine basis. A \emph{parallelotope} is a convex space whose
facets are pairwise parallel; it can be represented as a hypercube. See the
right side of fig.~\ref{fig:out_sp} for the two-dimensional case.

\paragraph{Convex forms}
\label{sec:plaus_forms}

We can consider mappings from a convex space to another, mappings that
preserve convex combinations. When the range is the real numbers, we can
speak of an \emph{affine form}, since such a convex mapping can be uniquely
extended to an affine form on the affine span of the convex space. This
kind of mappings are also defined for a mixture space, and
properties~\eqref{eq:conv1} and~\eqref{eq:conv2} are equivalent to say that
the mixture space is \emph{separated} or \emph{non-degenerate}, \viz, for
any pair of points there is a form having distinct values on them. In other
words, a convex space is a mixture space in which each pair of points can
be distinguished by a form \citep[\sect~3]{mongin2001}.

A surjective affine mapping from a convex space onto one of equal or lower
dimensionality can be called a (parallel) \emph{projection}. An injective
affine mapping from a convex space into one of higher dimensionality can be
called an (affine)
\emph{embedding}.

Affine forms from a convex space $\ySS$ to the interval $\clcl{0,1}$ are
especially important. We call them \emph{convex forms}. Convex
combinations of these can be naturally defined; they therefore constitute a
convex space, which can be given the name of \emph{convex-form space} (or
simply form space) of $\ySS$, denoted by $\yO{\ySS}$:
\begin{equation}
  \label{eq:outcomesp}
  \yO{\ySS} \defd
\set{\yom \colon \ySS \to \clcl{0,1} \st \text{$\yom$ is affine (or convex)}}.
\end{equation}
I avoid the name \enquote{dual space} because it risks to become overloaded
and easily confused with other notions of duality \citep[see
\eg][\sect~3.4]{gruenbaum1967_r2003}. The action of a convex form $v$ on
a point $a$ will be denoted by $v \inn a$ (confusion with affine forms on
affine spaces is not likely to arise); once an affine basis is chosen in
the convex space, this action can be written as matrix multiplication, as
for affine forms. The forms $\yoz \colon a \mapsto 0$ and
$\you \colon a \mapsto 1$ are called \emph{null-form} and \emph{unit-form}.

\begin{figure}[!t]
  \centering
  \includegraphics[width=10\columnwidth/10]{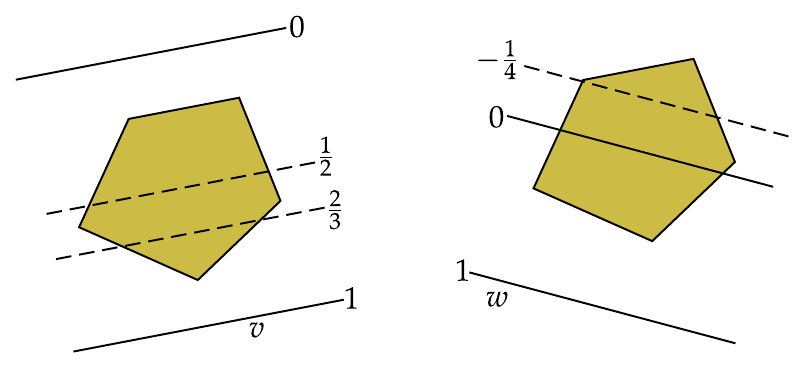}
  \caption{On the left, $v$ is a convex form for the five-sided
    convex space; two lines are indicated where the form has values $1/2$
    and $2/3$ on the space. On the right, $w$ \emph{cannot} be a
    convex form (although it is an affine form) because it assigns
    strictly negative values to some points of the convex space}
\label{fig:plausform}
\end{figure}

A non-constant convex form on a convex space can be geometrically
seen as a family of parallel hyperplanes (in the embedding affine space)
between two given ones that do not intersect the space's interior. On each
hyperplane the form has a constant value, with values zero and unity on the
utmost ones. These hyperplanes are also used to iconize the convex
form, a mark being put on the unit one; see fig.~\ref{fig:plausform}.

Convex forms allow us to give this definition: an \emph{exposed face} of a
convex space is a face on which a convex form has value $0$. Not all faces
are exposed faces; for example, the point $a_1$ and the similar point at
the other end of the curved boundary are non-exposed, zero-dimensional
faces. The presence of non-exposed faces has important consequences for
convex optimization, \ie\ the search for the extremum of a function over a
convex space (see references in the next section). 

The convex structure of a convex-form space is determined by that of its
respective convex space. Its affine span is the space of affine forms on
the affine span of the original convex space. This means, from what we said
about dual forms an bases in \sect~\ref{sec:affmaps}, that a form
space has one more dimension than the original convex space. A form
space is always a bi-cone with the null-form $\yoz$ and the unit-form
$\you$ as vertices; indeed, it is centro-symmetric with centre of symmetry
$(\yoz +\you)/2$. Its number of extreme points besides $\yoz$ and $\you$ is
determined by the structure of the faces of the original convex space (for
example, if the convex space is two-dimensional, the number of extreme
points of its form space is equal to $2m+2$, where $m$ is the
number of bounding directions of the convex space). See
fig.~\ref{fig:out_sp} for two two-dimensional examples. The convex-form
space of an $n$-simplex is an $(n+1)$-parallelotope (which has $2^{n+1}$
extreme points).

\subsection{References}
\label{sec:ref_convex}

Books on or touching convex spaces are Gr\"unbaum's
\citey{gruenbaum1967_r2003}, Valentine's \citey{valentine1964}, Alfsen's
\citey{alfsen1971}, Br{\o}ndsted's \citey{broendsted1983}, Eggleston
\citey{eggleston1958}. Studies and examples of the difference between
mixture and convex spaces are presented by Mongin \citey{mongin2001} and
Wakker \citey[\sect~VII.2]{wakker1988}. Other examples, axiomatizations,
and applications can be found in
\textcite{stone1949,hersteinetal1953,hausner1954,luceetal1971,luce1973,krantz1975,krantz1975b,vincke1980,holevo1980_t1982}.
See also
\textcite{gale1953,galeetal1968b,rockafellar1970_r1972,mcmullenetal1971,gruberetal1993,gruberetal1993b,schneider1993,webster1994,ewald1996,ball1997,bengtssonetal2006_r2017}
for related topics.

Convex optimization is a topic full of intriguing subtleties and is openly or
silently present in every branch of science. It has a vast literature,
sadly scattered between disciplines that do not talk with one another very
much. See
\textcite{fangetal1997,boydetal2004_r2009,borweinetal2000,berkovitz2002} as
possible starting points.

Infinite-dimensional convex spaces are less intuitive and require care in
their study. The studies of Klee and others
\citep{klee1948,klee1949,klee1949b,klee1950,klee1951,klee1951b,klee1951c,klee1953,klee1954,klee1955,klee1955b,klee1956b,klee1957,klee1958,klee1961,klee1963,klee1969,klee1969b,klee1969c,klee1969d,klee1969e,klee1971,klee1977,klee1980,burgeretal1996}
are very interesting and provide appropriate references.

\section{Generalizations: Grassmann spaces}
\label{sec:generalizations}

The operation of affine combination suggests several generalizations.

A question comes quite naturally to mind, for example: what if the
coefficients $(\ya_i)$ of an affine combination $\sum_i\ya_i a^i$ do not
sum up to one? In fact, in numeric applications it can be a nuisance to
make sure that the coefficients satisfy this requirement. It turns out that
an affine space can be seen as a special case of a more general space which
has various names in the literature; we call it \emph{Grassmann space}. A
Grassmann space is closed under an operation that looks
like~\eqref{eq:affine_sum}, with the exception that the coefficients $\ya$
and $\yb$ can assume arbitrary real values. Points in a Grassmann space and
in an affine space, however, differ: the former are equipped with a
\emph{weight}, which can be positive or negative. The operation
$\ya a + \yb b$ in a Grassmann space yields a point with weight $\ya+\yb$.
It is easy to guess that an affine space is like a Grassmann space where we
only consider points of unit weight. A remarkable consequence of this
generalization is that vectors and translations
(\sects~\ref{sec:geom_view}--\ref{sec:relation_an_geom}) turn out to be
\emph{points} with zero weight. Goldman \citey{goldman2002,goldman2000}
gives a brilliant introduction to these spaces.

A second question can naturally come to mind: could we consider affine
combinations not just of points, but also of straight lines, planes, and
analogous objects of higher dimensions? Also in this case the answer is
positive; in fact, we can define combinations with arbitrary coefficients.
The spaces where this is possible are again Grassmann spaces; Peano
\citey[\chap~I]{peano1888} gives an introduction to these generalized
combinations.

In fact, in a Grassmann space we can also define multiplicative operations
that combine points and lines, planes, and so on. This kind of spaces was
first consistently introduced by Grassmann
\citey{grassmann1844_r1878,grassmann1862}; Peano \citey{peano1888} also
gave a very accessible introduction to them. Unfortunately their subsequent
history -- which includes figures like Clifford \citey{clifford1878} and
Cartan \citey{cartan1923} -- has been very convoluted. Their multiplicative
operations have been developed by different groups of mathematicians in
ways that are inequivalent and, worst of all, overly complicated.
Interested readers can explore the approach by Barnabei, Brini, Rota, and
others \citep{barnabeietal1985,crapo2009,brinietal2011}; the approach by
Hestenes, Doran, Lasenby, Dorst, and others
\citep{hestenes1968,hestenesetal1984_r1987,dorstetal2002,li2008,dorstetal2011};
the approaches by Gunn \citey{gunn2011}, Browne \citey{browne2012},
Gonz\'alez Calvet \citey{gonzalezcalvet2016} -- and there are many others
out there \citep[see][remarks, \sect~1.4]{vargas2016}. Bengtsson and I
\citep{portamanaetal2017} hope to soon present an approach that makes
Grassmann spaces accessible to high-school students.

\section{An application: reference frames in classical mechanics}
\label{sec:ex_appl}

In the introduction I hinted at the fact that in classical
gal\-i\-le\-ian-\bd relativistic mechanics \emph{places} are often
represented by \enquote{position vectors} though they need not be modelled by
vectors at all; in fact some operations that we can do with vectors, \eg\
sums, do not have any physically meaningful counterpart for places. Places
can instead be modelled by a Euclidean space, which is a particular example
of affine space, one in which the additional notions of distance and angle
are defined. Velocities, accelerations, forces maintain their vectorial
character nevertheless. This is done as follows in the special case of
point-\bd mass mechanics:

We assume as primitives the notions of point-\bd mass, time, time lapse
(\ie\ a metric on the time manifold), and distance between any pair of
point masses at each time instant. We postulate that, at each time instant,
the net of distances among all point masses has a three-\bd dimensional
Euclidean character (\eg, theorems concerning triangles equalities and
triangle inequalities are satisfied). This net of distances determines
precise affine relations among the point masses; these relations are of
course variable with time like the distances themselves. The point masses
can therefore be made to span a three-\bd dimensional \emph{affine} space
at each time instant. The points of this affine space are what we call
\emph{places}, and each place is determined, in many equivalent ways, by an
affine combination of the point masses. For example, at an instant $t$ the
affine combination $a_1(t)/4 - 3 a_2(t)/4 + 6 a_3(t)/4$ determines a unique
place $b$ in terms of the point masses $a_1(t), a_2(t), a_3(t)$. Different
affine combinations can determine the same point: \eg, if $a_4(t) = 2
a_3(t) - a_2(t)$ at $t$, then $b$ is equivalently given by $a_1(t)/4 + 3
a_4(t)/4$. Note that, once the affine relations among the point masses are
given, we do not need a notion of absolute distance to determine $b$, nor
the ability to compare distances along unparallel directions; \ie\ we do
not need the Euclidean structure.

\begin{figure}[!t]
  \centering
  \includegraphics[width=10\columnwidth/10]{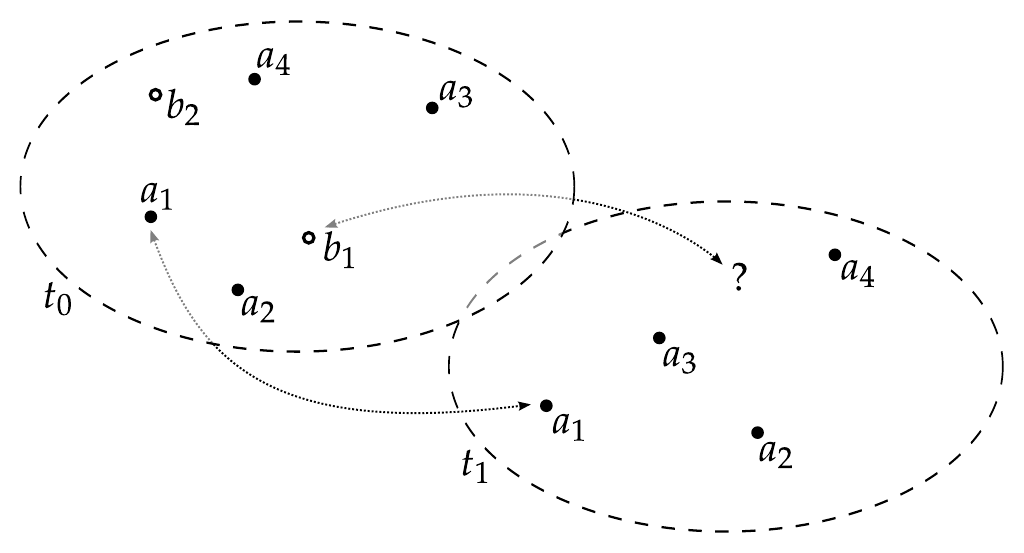}
  \caption{The Euclidean net of distances among the point masses $a_1, a_2,
    a_3, a_4$ determines an affine space at each time instant, \eg\ $t_0$
    and $t_1$. The point masses can be identified at each instant, but a
    generic place $b_1$ determined at $t_0$ by a particular affine
    combination of the point masses has no counterpart at $t_1$ because the
    affine relations among the point masses have changed.}
\label{fig:noident}
\end{figure}

At another time instant $t'$ the mutual distances and affine relations
between the point masses will be different; we may  have \eg\ $a_4(t') \ne 2
a_3(t') - a_2(t')$. So it does not make sense to try to identify at $t'$
the place $b$ that we defined at $t$: should it be given by $a_1(t')/4 - 3
a_2(t')/4 + 6 a_3(t')/4$?\ or by $a_1(t')/4 + 3 a_4(t')/4$?\ --- the two
combinations are inequivalent now. In other words, there is no
\emph{canonical} identification between the whole affine (and Euclidean)
spaces at two different instants of time. This also means that there is no
\enquote{absolute space}. See fig.~\ref{fig:noident}.

\begin{figure}[!t]
  \centering
 \includegraphics[width=10\columnwidth/10]{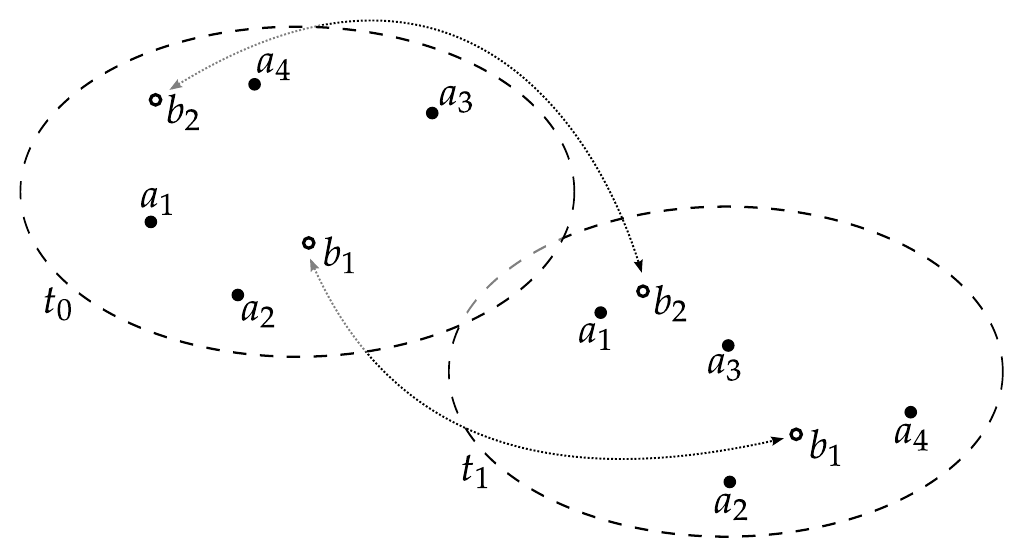}
 \caption{A \emph{frame of reference} is an arbitrary isomorphism between
   the places of the Euclidean-\bd affine spaces at any two times. With
   respect to the mapping above we can say, \eg, that the point mass $a_2$
   occupies the same place at $t_0$ and $t_1$, while the other point masses
   change place. Note, however, that the physical situation at $t_1$ (and
   $t_0$) in this figure and fig.~\ref{fig:noident} is exactly the same.}
\label{fig:frame}
\end{figure}

But absence of a canonical identification does not mean that no
identification at all is possible. A \emph{frame of reference} is a
particular, \emph{arbitrary} identification of the places of the affine
spaces at any two instants of time, respecting the affine and Euclidean
structure; \ie, it is a mapping, defined for any two instants $t$ and $t'$,
\begin{gather}
  \label{eq:frame_map}
  \yF{t',t} \colon \afs_{t'} \to \afs_{t},
\end{gather}
between the Euclidean-\bd affine spaces $\afs_{t}'$, $\afs_{t}$ spanned by
the point masses at those two instants, that preserves distances. It
therefore preserves affine combinations:
\begin{equation}
  \label{eq:frame_pres_aff}
  \yF{t',t}(\ya a' + \yb b') = \ya \yF{t',t}(a') + \yb \yF{t',t}(b').
\end{equation}
A frame of reference allows us to say that a particular place at time $t$
is the \enquote{same} as some place at time $t'$, so that we can use only one
affine space for all times and we can say that a particular point mass
\enquote{moved} from a place at $t$ to another place at $t'$. See
fig.~\ref{fig:frame}.

The velocity of a point mass' motion at the instant $t_0$ in a particular
frame $\yF{}$ is defined as
\begin{equation}
  \label{eq:veloc_aff}
  \yv(t_0) \defd 
\lim_{t \to t_0} \frac{\yF{t,t_0}[p(t)] - p(t_0)}{t-t_0},
\end{equation}
$p(t)$ being the place occupied by the point mass at time $t$. The argument
of the limit is, for each $t$, the \enquote{difference} between two points in the
affine space associated to the instant $t_0$: it is namely a translation,
as discussed in \sect~\ref{sec:geom_view}, and therefore a vector. The
limit is hence a vector, too. In this way we obtain the vectorial character
of velocities, accelerations, and in a similar way of forces, without the
need to model places as vectors. Note again that only the affine structure
of space enters in the expression above, not the Euclidean one (but we have
a metric on the one-dimensional manifold that models time, as implied by
the denominator of the fraction).

This way of modelling space in classical mechanics is based upon and
combines the works of Noll \citey{noll1959,noll1967}, Truesdell
\citey{truesdell1977_r1991}, and Zanstra
\citey{zanstra1922,zanstra1923,zanstra1924,zanstra1946}. Apart from
mathematical economy, it has the pedagogic advantage of presenting
gal\-i\-le\-ian-\bd relativistic mechanics in a fashion closer to that of
general relativity: in general relativity the set of events is a manifold
that cannot be modelled as a (four-dimensional) vector space. Only
(four-)velocities, accelerations, momenta have a vectorial character.

\setlength{\intextsep}{0.5ex}

\ifpublic
\begin{acknowledgements}
  \ldots to Ingemar Bengtsson for support, for many, always insightful convex and
  non-convex discussions, for references, and for pointing out deficiencies
  in previous versions of this note; any deficiencies that \sout{may}
  remain are my fault. To Mari \amp\ Miri for continuous encouragement and
  affection. To Buster Keaton and Saitama for filling life with awe and
  inspiration. To the developers and maintainers of \LaTeX, Emacs, AUC\TeX,
  Open Science Framework, biorXiv, Hal archives, Python, Inkscape, Sci-Hub
  for making a free, unfiltered, and unmoderated scientific exchange possible.
\sourceatright{\autanet}
\end{acknowledgements}
\fi



\defbibnote{prenote}{{\footnotesize (\enquote{de $X$} is listed under D,
    \enquote{van $X$} under V, and so on, regardless of national
    conventions.)\par}}

\printbibliography[prenote=prenote
]

\end{document}